\documentclass[preprint]{ptephy_v1}

\preprintnumber{XXXX-XXXX} 

\newcommand{\gsim}{{ \hspace{0.3em}\raisebox{0.4ex}{$>$}\hspace{-0.75em}\raisebox{-.7ex}{$\sim$}\hspace{0.3em} }}

\begin{document}

\title{Thermal surface neutralization of Fr ions with metal foils for magneto-optical trapping of radioisotopes}


\author[1,2,*]{H.~Kawamura}
\affil{Frontier Research Institute for Interdisciplinary Sciences,  Tohoku University, Miyagi 980-8578, Japan\email{kawamura@cyric.tohoku.ac.jp}}

\author{T.~Aoki}
\affil{Cyclotron and Radioisotope Center, Tohoku University, Miyagi 980-8578, Japan}

\author[2]{K.~Harada}
\author[1,2]{T.~Inoue\footnote{Present address: RIKEN Nishina Center for Accelerator-Based Science, 2-1 Hirosawa, Wako, Saitama 351-0198, Japan}}
\author[2]{S.~Ito}
\author[2]{K.~Kato}
\author{L.~K\"{o}hler}
\affil{Technische Universit\"{a}t Darmstadt, Darmstadt 64289, Germany}
\author[2]{K.~Sakamoto}
\author[2]{A.~Uchiyama}

\author{Y.~Sakemi}
\affil{Center for Nuclear Study, the University of Tokyo, Tokyo 113-0033, Japan}




\begin{abstract}%
We investigate neutralization processes (especially thermal surface neutralization), which are required for the magneto-optical trapping of radioactive atoms. 
A variety of neutralization methods are first summarized: 
neutral beam injection for fusion reactors, neutral atom implantation in semiconductor processing, and the production of radioactive neutral atoms in accelerators. 
We focus on thermal surface neutralization, which produces neutral atoms in the thermal energy range for laser cooling. 
The experiments were carried out with yttrium, gadolinium, and zirconium foils to neutralize francium and rubidium ions for magneto-optical trapping. 
The results reconfirm that yttrium foil is a good neutralizer (i.e., it has a neutral release efficiency $>65\%$). 
In addition, the release fraction when using yttrium foil exceeds 75\% at 1350 K, which is greater than the release fraction for the other foils. 
This reconfirmation is important because few previous studies have focused on thermal surface neutralization. 
Moreover, the results show that the neutralization efficiency is strongly influenced by the experimental process itself. 
\end{abstract}

\subjectindex{H00, I90}
\maketitle

\section{Introduction}
Laser cooling and trapping of atoms has led to an evolution in our understanding of atomic physics. 
For example, it has led to a new generation of atomic clocks and to the realization of Bose-Einstein condensation~\cite{MetcalfStraten}. 
Recently, magneto-optical traps (MOTs) have been used to trap short-lived radioactive alkali elements~\cite{0954-3899-36-3-033101, doi:10.1146/annurev.nucl.47.1.429}, and studies are being planned to use MOTs to trap radioactive Fr~\cite{Zhang2016, doi:10.1063/1.4896609, PhysRevX.2.041009, 1742-6596-302-1-012051}. 
Because Fr is the heaviest alkali metal, it is a good candidate to search for the electron electric dipole moment and to measure atomic parity nonconservation~\cite{0954-3899-36-3-033101}. 
When radioisotopes are produced within an accelerator, they are often ionized to extract more particles~\cite{Stora:1693046}. 
However, for magneto-optical trapping, alkali elements must be neutral atoms. 
Therefore, in general, the radioactive ion is converted into a neutral atom before being trapped. 
This paper starts by discussing candidates for a neutralization method for MOTs. 
Next, our experimental results are reported for thermal surface neutralization, which is one of the candidates. 
This process uses a simple mechanism to produce neutral atoms in the thermal energy range for laser cooling. 

\subsection{Neutralization methods}
Conversion from ions to neutral atoms is required in various situations. 
For example, neutral beam injection is used as one component of the heating system in fusion reactors~\cite{DALLAPALMA20131020}. 
Neutral beam injection uses charge-exchange reactions with gas atoms in a gas cell to convert a high-energy ion beam into a beam of high-energy neutral particles. 
This method is also used for the surface cleaning and depth profiling of semiconductors~\cite{1347-4065-33-11R-6325}. 
However, when the incident beam energy is less than 1 eV available for laser cooling, the output cannot be part of the beam because scattering in the gas is considerable. 

A radioactive neutral atomic beam can be produced by using a charge-exchange reaction with alkali vapor~\cite{LEVY2002253, doi:10.1063/1.3271037}. 
For alkali elements, the energy dependence of the cross section for charge exchange with the same type of atom as in the incident beam differs from that with a different type of atom~\cite{PhysRev.138.A937}. 
A Fr ion beam cannot be neutralized by using a Fr gas because Fr is a short-lived radioactive element. 
The cross section of the reaction with different elements decreases with decreasing beam energy. 
For example, the charge-exchange cross section between Rb vapor and a 5 keV Fr beam is $(9\pm3)\times 10^{-15}\ \mathrm{cm}^2$~\cite{doi:10.1063/1.3271037}. 
For Rb vapor and a Cs beam, the cross section is an order of magnitude less for an ion velocity reduced by an order of magnitude~\cite{PhysRev.138.A937}. 
Thus, in view of the cross section, the charge-exchange reaction is unsuitable to neutralize Fr for MOTs. 

Industrial-grade semiconductor devices use a process in which ions colliding with a solid surface scatter and form neutral atoms~\cite{1347-4065-40-10A-L997}. 
Neutral atom injection can be used to implement damage-free surface processes (etching, beam deposition, etc.). 
This process is accomplished by resonant neutralization~\cite{ZIMNY1990333} or Auger neutralization~\cite{PhysRevB.58.13991}. 
The energy of the output atoms is mostly in the hyperthermal region ( 1--100 eV), which is too high for laser cooling and trapping. 
Originally, the colliding neutralization method and the neutral beam injection method for reactors were developed to produce high-energy neutral atoms, which cannot be achieved using the common heating method. 
High-energy neutral atoms are unsuitable for laser cooling, which requires atoms in the thermal energy range. 

Radioactive neutral atoms from a production target were directly laser cooled and trapped without undergoing neutralization~\cite{PhysRevLett.72.3791, PhysRevLett.93.102501}. 
This method has already produced successful results and is projected to be one of the best candidates for MOTs. 
However, the method is problematic because the trapping apparatus must be placed near the production target, which is at the end of the primary beam line. 
For example, at Lawrence Berkeley Laboratory~\cite{PhysRevLett.72.3791}, the trapping apparatus was placed roughly 2 m from the production target with over 1 m of radiation shielding between them. 
A change in the primary beam line and installation of the radiation shielding might require modification of the entire facility, which could not easily be carried out. 

Another way of obtaining radioactive neutral atom beams is to desorb alkali atoms from metal targets by low-intensity nonresonant optical irradiation. 
This method is preferable for efficient trapping because it does not degrade an ultrahigh vacuum, unlike thermal desorption. 
This neutralization method has entered the discussion because Coppolaro et al. used a photographic flash light in 2014 to desorb Fr from a Y foil~\cite{doi:10.1063/1.4896609}. 

A more general method to trap radioactive atoms~\cite{PhysRevLett.72.3795, PhysRevLett.76.3522, PhysRevLett.79.375} is as follows. 
Radioisotopes are extracted from the production target as ions, which are relatively easy to handle. 
Unlike neutral atoms, charged particles are subject to Coulomb and Lorentz interactions with electric and magnetic fields, respectively, which facilitates extraction, acceleration, and focus. 
The isotopes are transported as an ion beam, neutralized, and then trapped. 
The neutralization method is based on thermal surface neutralization using a metal target with a small work function. 
This method can convert a keV ion beam into thermal atoms (a few hundred of degrees Celsius, or $\sim0.1$ eV) suitable for laser cooling. 

Thermal surface neutralization has also been used for other experiments. 
An orthotropic source can produce neutral atoms from an oven, including radioactive isotopes~\cite{doi:10.1063/1.1146804}. 
This technique has been used to trap Fr isolated from an accelerator in a MOT~\cite{PhysRevLett.79.994}. 
Also, the atomic beam resonance method has been used to study atomic structure by producing radioactive atomic beams via thermal surface neutralization~\cite{TOUCHARD1981329,  PINARD20053, Nagae2005580}. 
In order to efficiently perform such experiments, experimental apparatuses for generating atomic beams with small angular divergence have been developed~\cite{Francois2018}. 
Although thermal surface neutralization has been used for several experiments, this process has not developed significantly. 
A better understanding of the neutralization process is expected to extend its range of application. With this motivation, we investigate in the present work thermal surface neutralization. 

We thus focus herein on thermal surface neutralization as a conversion process for the MOT of radioactive atoms. 
To determine the emitted components of particles implanted in the target, we monitor not only the neutral output but also the charged output. 
The experiment and its results are discussed in the following sections. 
The results of neutralization of Fr using Y, Gd, and Zr confirm that the neutralization efficiency roughly follows the Saha-Langmuir equation. 
In addition, Rb neutralization by Y indicates that the neutralization efficiency depends on experimental conditions and can easily vary. 

\subsection{Thermal surface neutralization}

In this paper, ``thermal surface neutralization'' refers to the phenomenon in which ions incident on a metal surface are released as neutral atoms on heating the surface. 
The use of a high-temperature metal surface is a well-known method to produce an ion beam~\cite{Stora:1693046, Traykov20084478}. 
This process is described by the Saha-Langmuir equation~\cite{Langmuir61} 
\begin{equation}\label{Eq:SahaLangmuir}
\eta = \frac{n_{\mathrm{atom}}}{n_{\mathrm{ion}}+n_{\mathrm{atom}}} = \frac{1}{n_{\mathrm{ion}}/n_{\mathrm{atom}}+1}
\end{equation}
\begin{equation}
\textrm{with}\ \frac{n_{\mathrm{ion}}}{n_{\mathrm{atom}}} = \frac{g_{\mathrm{ion}} }{g_{\mathrm{atom}} } \exp \left( \frac{E_{\mathrm{WF}}-E_{\mathrm{IP}} }{k_{\mathrm{B}} T } \right), 
\end{equation}
where $n_{\mathrm{ion}}$ is the ion density on the metal surface, $n_{\mathrm{atom}}$ is the atom density, and $g_{\mathrm{ion}}$ and $g_{\mathrm{atom}}$ are the statistical weights of the ion and atom, respectively. 
In particular, $g_{\mathrm{ion}}=1$ and $g_{\mathrm{atom}}=2$ for alkali elements. 
The energy $E_{\mathrm{WF}}$ is the work function of the metal surface, $E_{\mathrm{IP}}$ is the ionization potential of the atom (4.07 eV for Fr), $k_{\mathrm{B}}$ is the Boltzmann constant, and $T$ is the absolute temperature of the metal surface. 
According to this equation, the ion density increases for $E_{\mathrm{WF}} > E_{\mathrm{IP}}$, and the atom density increases when $E_{\mathrm{WF}} < E_{\mathrm{IP}}$. 
Note that, depending on the situation, this process can also produce negative ions~\cite{KAWANO19831}. 
As summarized in Table \ref{tab:studies}, few reports focus on the neutralization process of this surface reaction. 
Three efficiencies characterize the neutralization process: the total release efficiency $\varepsilon_{\mathrm{all}}$ indicates the number of neutral or charged particles extracted per ion injected into the target, the neutral release efficiency $\varepsilon_{\mathrm{atom}}$ indicates the number of neutral atoms extracted per injected ion, and the neutral ratio $\eta$ indicates the percent of emitted particles that are neutral atoms. 
Most previous studies measured only one of these efficiencies. 
Besides, they measured the efficiency of trapping neutral atoms with metal foil at different temperatures and with different implant energies~\cite{320869693d2e40758a6e68123918d091, Kawamura2015}. 

Yttrium has been used to neutralize radioactive alkali ions because the work function $E_{\mathrm{WF}}$ of Y is reportedly much smaller than the ionization potential $E_{\mathrm{IP}}$ of alkali atoms. 
Melconian et al. ~\cite{Melconian200593} compared different materials (V, Fe, Ni, Y, Zr, Nb, Mo, Hf, Ta, W, Re, C, and Pt) as neutralizers of ${}^{37}\mathrm{K}$ in 2005. 
However, they did not monitor whether the released particles were atoms or ions. 
Guckert~\cite{Guckert1998} studied the release of ${}^{82}\mathrm{Rb}$ from Y, Hf, Ta, Mo, and Pt foils as neutralizers in 1998. 
He estimated the fraction released as neutrals by monitoring the release with and without a negative suppression voltage on the neutralizer. 
Few reports state whether the released particles are neutral atoms or ions because it is difficult to detect ions and atoms with a single method. 
The experiment reported herein shows that the released atoms are separated from the ions. 

\begin{table}[!h]
\caption{Summary of neutralization studies using Fr, Rb, K, and Al beams incident on Y foil. }
\label{tab:studies}
\centering
\begin{tabular}{c c c c c c}
Atom & Y temperature & Total release                & Neutral release                & Neutral ratio & Reference  \\ 
     & $T$ (K)            &  $\varepsilon_{\mathrm{all}}$ (\%) & $\varepsilon_{\mathrm{atom}}$ (\%) & $\eta$ (\%)  &      \\ \hline \hline
Fr-210\&211& $\sim1350$ & $\gsim75$ & $\gsim 65$ & $\gsim 85$ & This work \\ \hline
Fr & 1200 & $\sim47$ &  &  & \cite{LIPSKI200449} \\ \hline
Fr-210 & 1080 &  & 90  &  & \cite{PhysRevA.78.063415} \\ \hline
Fr & $\sim900$ &  &  &  $\sim30$ &  \cite{KawamuraCAR14} \\ \hline
Rb-82 & 1313 & 40 & 40 & 100 & \cite{Guckert1998} \\ \hline
Rb-82 & 1023 &    & 30 &   & \cite{Guckert1998} \\ \hline
K-37 & $\sim1340$ & $\sim65$ &  &  & \cite{Melconian200593} \\ \hline
Al-30 & ${}^{\mathrm{Room}}_{\mathrm{temperature}}$ &  &  &  $\sim94$ & \cite{Nagae2005580}  
\end{tabular}
\end{table}

\subsection{Effective work function}

Note that the work function depends on the condition of the surface. 
Different work functions are often given for Y: 
3.1 eV~\cite{LIPSKI200449}, 
3.2 eV~\cite{TOUCHARD1981329}, or
3.4 eV~\cite{doi:10.1063/1.1146804}. 
Eastman~\cite{PhysRevB.2.1} reported that the Y work function is $(3.1\pm0.15)$  eV. 
He measured the photoelectric work function of a clean Y surface evaporated in high vacuum ($10^{-10}$ to $10^{-8}$ Torr). 
In general, a metal foil is used for thermal surface neutralization. 
Assembling or welding is required to implement the foil into the apparatus. 
If this operation is conducted in an ambient atmosphere, the foil will oxidize, which causes the work function to vary. 
For example, the work function $E_{\mathrm{WF}}$ of yttrium oxide on a tungsten cathode, which is used in electron guns, is estimated to be 2.0 eV~\cite{5271574} less than that of pure Y. 
Furthermore, oxidized, porous tungsten has a larger work function, which leads to an increased surface ionization efficiency~\cite{doi:10.1063/1.1720217}. 
The work function $E_{\mathrm{WF}}$ for clean tungsten is about 4.6 eV but rises to over 6.0 eV for an oxygenated tungsten surface. 
The ion-current density without oxygen flow is approximately 0.1 $\mu\mathrm{A/cm}^2$, whereas that with oxygen flow exceeds 1.3 $\mu\mathrm{A/cm}^2$. 
The neutralizer for trapping must undergo some processes that can vary the surface condition, 
including heating~\cite{KAWANO1986195}, beam sputtering~\cite{BRISON20071467}, and beam deposition~\cite{PhysRevLett.40.574}. 
Because the release fraction is not always 100\%, the surface condition will change between every experimental sequence because of the deposition. 
Therefore, the thermal surface neutralization must in reality be discussed in terms of an effective work function rather than an ideal work function. 

\section{Experiment}

The experiments with radioactive Fr and stable Rb were performed with Y, Gd, and Zr foils. 
These materials have proven records as targets to convert alkali metal ions into neutral atoms~\cite{LIPSKI200449, PhysRevLett.86.2967, PhysRevA.97.042507}. 

\subsection{Neutralization of Fr by Y}

This experiment was carried out at the Fr-dedicated beam line at the Cyclotron and Radioisotope Center at Tohoku University~\cite{KAWAMURA2013582}. 
Fr is produced via a nuclear fusion evaporation reaction between an Au target and an ${}^{18}\mathrm{O}$ beam from an AVF 930 cyclotron. 
The Fr produced is ionized on the hot gold surface ($\sim1000\ ^{\circ}\mathrm{C}$) and extracted and transported by electrostatic fields. 
The Fr ion beam was irradiated on a 25-$\mu$m-thick, 10 mm $\times$ 10 mm Y neutralizer foil, and neutralized Fr atoms were detected. 

Kawamura et al.~\cite{KawamuraCAR14} revealed low-efficiency neutralization ($\eta\sim30\%$) with Y foil, which contradicts the conventional understanding; 
the efficiency $\eta$ should be almost unity according to the Saha-Langmuir equation (\ref{Eq:SahaLangmuir})~\cite{doi:10.1063/1.4896609, TOUCHARD1981329, PhysRevA.78.063415}. 
However, Kawamura et al.'s study~\cite{KawamuraCAR14} failed to account for the fact that the given Y foil had been used for many experiments. 
Before measurement of the neutralization efficiency, this Y foil was used to neutralize ions for the MOT of neutral atoms originating from the ion beam. 
In other words, this foil had previously been significantly irradiated by Rb, Fr, and the background beam, heated many times, and exposed to the ambient atmosphere when the apparatus was reassembled. 
In addition, the Fr ion beam had low purity, which could further influence the surface condition, although a Wien filter~\cite{doi:10.1063/1.4935013} was implemented to purify the beam after that measurement. 
The beam purity, which is defined as the ratio of Fr beam intensity to total beam intensity, was approximately $10^{-6}$ for the previous neutralization experiment~\cite{KawamuraCAR14}. 
To avoid the difficulties of this previous study, the goal in the present experiment is to measure the neutralization efficiency of Y foil under the best conditions 
(minimization of beam irradiation not related to the measurement, elimination of unnecessary heating, and no exposure of the apparatus to the ambient atmosphere). 

\begin{figure}[!h]
\centering\includegraphics[width=150mm]{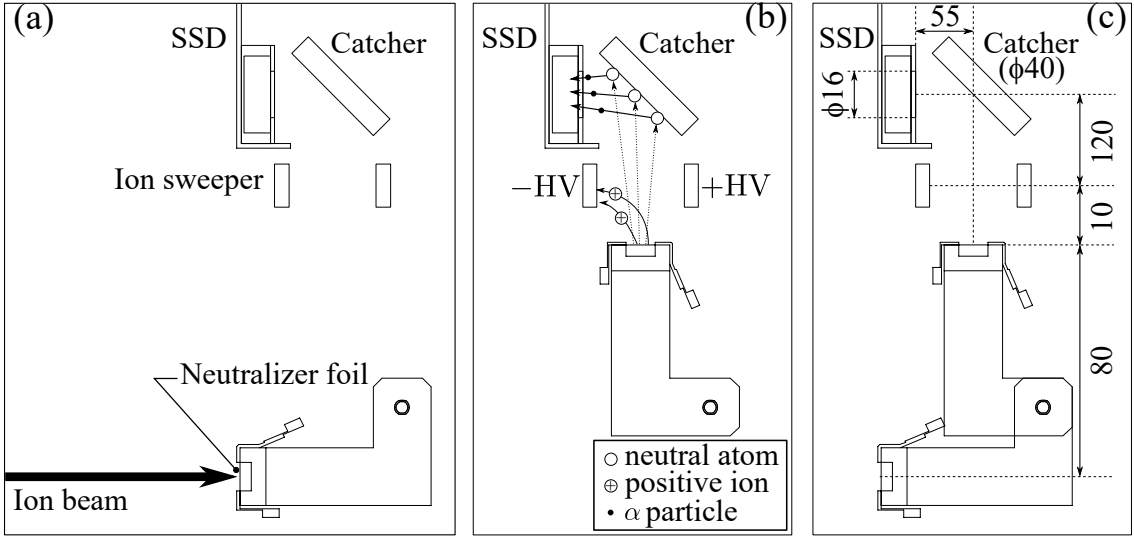}
\caption{
Setup of the measurement of Fr neutralization with Y foil. 
The effective area of the target is 10 mm $\times$ 10 mm. 
(a) When the target is positioned downward, the ion beam accumulates on the foil. 
(b) When the target is positioned upward and the ion sweeper is working, the SSD detects only $\alpha$ particles originating from the neutralized Fr atoms. 
(c) Schematic plan of the setup with dimensions in mm. 
In this configuration, $\alpha$ rays emitted from the Y foil do not reach the SSD directly. 
Because the thermal radiation emitted by the foil was detected by the SSD, causing noise, no data acquired during heating were included in the analysis. 
}
\label{Fig:SetupFr}
\end{figure}

\begin{figure}[!h]
\centering\includegraphics[width=110mm]{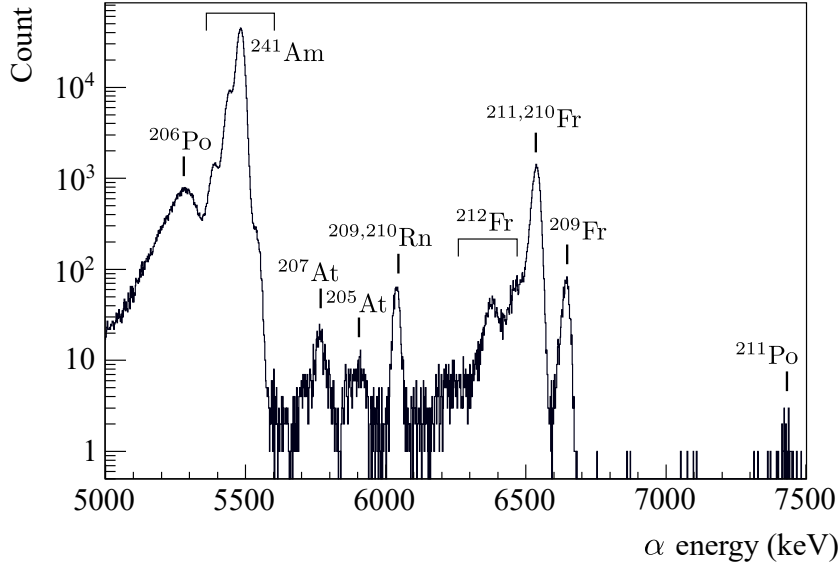}
\caption{Energy spectrum of $\alpha$ particles detected by the SSD. 
${}^{241}\mathrm{Am}$ was used for energy calibration. 
The other particles were products of the fusion reaction or their progeny nuclides. 
}
\label{Fig:sumevent_noiseremove}
\end{figure}

The experimental procedure was as follows. 
First, a 3 keV Fr beam irradiated a neutralizer foil for 200 s, as shown in Fig.~\ref{Fig:SetupFr}(a). 
Because Fr is radioactive (the half-life of ${}^{210}\mathrm{Fr}$ is 3.18 min and that of ${}^{211}\mathrm{Fr}$ is 3.10 min), the number of accumulated particles is not proportional to the accumulation time. 
If the saturated number of particles for infinite time is 100\%, 200 s of accumulation yields over 50\%. 
We determined that accumulation for 200 s leads to a reasonable measurement; furthermore, excessively long measurement might cause systematic errors. 
The beam current, which was purified through the Wien filter, was typically $(2.8\pm0.2)$ pA, which included $(1.7\pm0.3)\times10^{3}$ ${}^{210,211}\mathrm{Fr}$ ions per second. 
Next, the neutralizer target was turned upward, as shown in Fig.~\ref{Fig:SetupFr}(b), and heated to release the Fr. 
The sequence for target heating was as follows: the electric current increases from 0 A to the desired current value over the initial 2 s, following which the current is held constant for the next 10 s. 
This experientially determined sequence was found to be optimal to magneto-optically trap the neutralized atoms. 
We hypothesized that most atoms are emitted over the first few seconds, and long heating times generate numerous background particles, which degrade the trapping efficiency. 
After heating, the target was maintained in the upward state for over 290 s to ensure adequate time to conduct $\alpha$-ray spectroscopy. 
Next, the target was turned downward, as in Fig.~\ref{Fig:SetupFr}(a), and once again irradiated by the ion beam. 
The charged particles released from the neutralizer were swept out by a sweeper field, which was generated by a pair of electrodes above the neutralizer. 
The effective area of the sweeper electrode was 20 mm $\times$ 30 mm, the gap between electrodes was 20 mm, and the applied voltage was $\pm$1000 V. 
The neutral particles were gathered by a catcher plate placed above the sweeper. 
The catcher plate consisted of a copper disk 40 mm in diameter. 
The radioactive atoms decayed on the catcher and emitted $\alpha$ particles at the rate determined by the decay constant of each radioisotope 
(for ${}^{210}\mathrm{Fr}$, the decay constant $\lambda=3.63\times10^{-3}$ $\mathrm{s}^{-1}$ and the $\alpha$-decay branching ratio $I_{\alpha}=0.71$; for ${}^{211}\mathrm{Fr}$, $\lambda=3.73\times10^{-3}$ $\mathrm{s}^{-1}$ and $I_{\alpha}=0.87$). 
Particle detection and identification was carried out by $\alpha$-ray spectroscopy using a silicon semiconductor detector (SSD). 
The SSD could not resolve the $\alpha$ energy of ${}^{210}\mathrm{Fr}$ (6.545 MeV) and ${}^{211}\mathrm{Fr}$ (6.537 MeV), so these two isotopes were counted together. 
The energy resolution of the SSD was 27 keV in the energy range of interest. 
The stacked $\alpha$-energy spectrum is shown in Fig.~\ref{Fig:sumevent_noiseremove}. 
In the previous study~\cite{KawamuraCAR14}, ions could display unexpected behavior because of the complex sweeper field between an anode electrode and a grounded vacuum chamber, which corresponds to the cathode. 
In the present work, the anode and cathode were separated from the chamber. 
Because positive ions were gathered onto the cathode, $\alpha$ particles from the cathode could not arrive at the SSD. 
Test experiments with Rb and Re foils, which have a high ionization efficiency, demonstrated that ions cannot pass through the sweeper field. 
The neutralizer foil was resistively heated by running an electric current through the foil. 
The temperature of the foil was determined based on a correspondence table that mapped the applied current to the temperature, which was measured by using a radiation thermometer. 
The measured temperature was roughly 1200 K for an electric current of 8 A, and roughly 1500 K for 12 A. 

The efficiencies related to the neutralization process are defined as follows. 
The number of Fr remaining on the Y foil after beam irradiation is $N_{\mathrm{acc}}$. 
By heating the foil, Fr ions and neutral atoms are released with efficiencies $\varepsilon_{\mathrm{ion}}$ and $\varepsilon_{\mathrm{atom}}$, respectively. 
Altogether, the release efficiency for all Fr is $\varepsilon_{\mathrm{all}} = \varepsilon_{\mathrm{ion}} + \varepsilon_{\mathrm{atom}}$. 
The ratio of arrival at the catcher is $\Omega_{\mathrm{c}}$, and $\Omega_{\mathrm{s}}$ is the efficiency with which the $\alpha$ particles from the catcher are detected by the SSD. 
The total count $N_{\mathrm{all}}$ at the SSD over $t$ seconds and in the absence of the sweeper field is
\begin{equation}
N_{\mathrm{all}} = \varepsilon_{\mathrm{all}} \Omega_{\mathrm{c}} \Omega_{\mathrm{s}} N_{\mathrm{acc}} (1-e^{\lambda t}), 
\end{equation}
where $\lambda$ is the decay constant of the given radioisotope. 
Because only neutral atoms are detected on the sweeper field, 
\begin{equation}
N_{\mathrm{atom}} = \varepsilon_{\mathrm{atom}} \Omega_{\mathrm{c}} \Omega_{\mathrm{s}} N_{\mathrm{acc}} (1-e^{\lambda t}). 
\end{equation}
By determining the ratio of these numbers, we derive the fraction of released Fr that is neutral
\begin{equation}\label{Eq:etaDef}
\eta = \frac{N_{\mathrm{atom}}}{N_{\mathrm{all}}} = \frac{\varepsilon_{\mathrm{atom}} }{\varepsilon_{\mathrm{ion}} + \varepsilon_{\mathrm{atom}} } 
\end{equation}
: The quantity $\eta$ is defined as the neutral ratio. 

\begin{figure}[!h]
\centering\includegraphics[width=80mm]{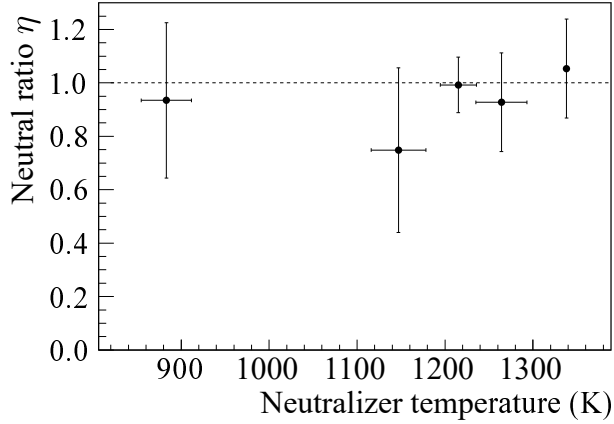}
\caption{
Neutral ratio $\eta$ of Fr as a function of the temperature of the Y foil. 
The vertical error bars give the statistical error. 
The horizontal error bars are estimated using the current-temperature correspondence table. 
The temperature error near 1340 K is small because the current-temperature correlation is high at this temperature. 
}
\label{Fig:eta_Y}
\end{figure}

Figure \ref{Fig:eta_Y} shows the neutral ratio $\eta$ measured at different temperatures. 
In the present measurement range, $\eta$ is always coincident with 100\% within the error. 
Figure \ref{Fig:eta_Y} shows the results for ${}^{210}\mathrm{Fr}$ and ${}^{211}\mathrm{Fr}$. 

\begin{figure}[!h]
\centering\includegraphics[width=80mm]{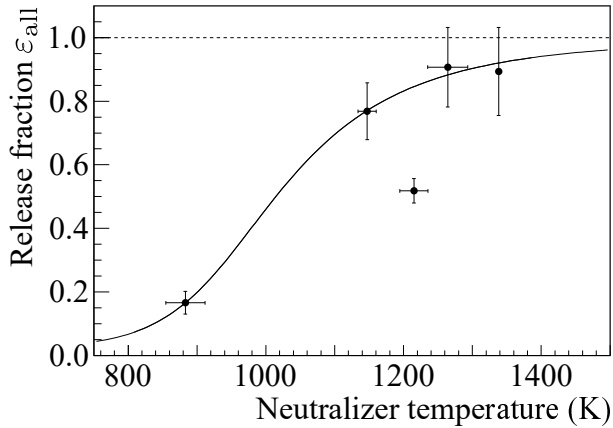}
\caption{
Total release efficiency $\varepsilon_{\mathrm{all}}$ of Fr as a function of the temperature of the Y foil. 
The solid curve is from Eq.~(\ref{Eq:releasefrac}) and serves to guide the eye. 
}
\label{Fig:eps_all_Y}
\end{figure}

The temperature dependence of the total release efficiency $\varepsilon_{\mathrm{all}}$ was derived. 
The results of the earlier studies~\cite{PhysRevA.78.063415} were used to determine $\varepsilon_{\mathrm{all}}$ and the product $\Omega_{\mathrm{c}} \Omega_{\mathrm{s}}$. 
The diffusion coefficient of Fr in Y at 1000 K was measured to be $\tau_{1000}=(4.1\pm2.5)$  s~\cite{PhysRevA.78.063415}. 
For $N_{\mathrm{acc}}$ ions all implanted at the same distance from the surface at time $t=0$, the diffusion equation has an analytical solution. 
For an implantation distribution with a finite extent, the function can differ. 
For instance, Melconian et al.~\cite{Melconian200593} considered a distribution modeled by a Gaussian multiplied by a linear term, with a characteristic implantation depth. 
Under such an assumption, Mauro et al.~\cite{PhysRevA.78.063415} obtained the following function for flux released from a solid surface: 
\begin{equation}
F(t) = \frac{ N_{\mathrm{acc}} }{ 2 } \frac{ 1 }{ \tau_{\mathrm{d}} }\frac{ 1 }{ (1+t/\tau_{\mathrm{d}})^{3/2} } \exp(-\lambda t), 
\end{equation}
so when the temperature is constant (1000 K) for 10 s, the total release efficiency is 
\begin{equation}
\varepsilon_{\mathrm{all}} (T=1000\,\mathrm{K}) = \frac{\int_{0}^{10} F_{1000\,\mathrm{K}}(t)dt }{ N_{\mathrm{acc}}} = 0.45^{+0.17}_{-0.09}. 
\end{equation}
By using the equation for the release fraction~\cite{Melconian200593}, we obtain
\begin{equation}\label{Eq:releasefrac}
\varepsilon_{\mathrm{all}} (T)= \exp \left( -\frac{1}{\sqrt{\alpha} } \right) \frac{\sinh(f/\sqrt{\alpha}) }{f/\sqrt{\alpha} }, 
\end{equation}
where the parameter $\alpha = (\tau/d^2) D_0 \exp(-E_{\mathrm{a}} / k_{\mathrm{B}}T)$~\cite{Melconian200593}. 
The product $\Omega_{\mathrm{c}} \Omega_{\mathrm{s}}$ was modified because $\varepsilon_{\mathrm{all}}=0.45$. 
Figure~\ref{Fig:eps_all_Y}, which shows the temperature dependence of $\varepsilon_{\mathrm{all}}$, uses this product $\Omega_{\mathrm{c}} \Omega_{\mathrm{s}}$, although the vertical error bars include only statistical errors from the measurement, not the error in $\Omega_{\mathrm{c}} \Omega_{\mathrm{s}}$. 
We confirmed that the data are qualitatively described by Eq.~(\ref{Eq:releasefrac}). 
However, significant parameters in Eq.~(\ref{Eq:releasefrac}) are not derived from this result because the measured data and the quoted diffusion coefficient include large errors. 
The deviation of the data near 1200 K seems to be due to a fluctuation in ion-beam intensity. 
It is unlikely that the release fraction decreases at a certain temperature due to any intrinsic physical phenomenon. 
The accumulated number $N_{\mathrm{acc}}$ is estimated from the intensity measured by a beam monitor while the beam is not accumulating on the foil. 
In addition, the beam intensity sometimes decreased differentially because it included many fluctuation factors, such as the intensity of the oxygen beam from the cyclotron, the condition of the gold target, and the stability of the electrostatic transportation. 
The deviation near 1200 K is therefore attributed to a fluctuation in beam intensity that the monitor did not detect. 

\subsection{Neutralization of Rb by Y}

The present study obtains a different neutralization efficiency than our previous study~\cite{KawamuraCAR14}. 
When the temperature of the Y foil is approximately 900 K, the neutral ratio $\eta$ exceeds 65\% in the present work, whereas it was roughly 30\% in the previous work~\cite{KawamuraCAR14}. 
Actually, earlier studies also obtained different efficiencies, as shown in Table \ref{tab:studies}. 
The present measurement was performed to elucidate whether the experimental process itself influences the neutralization efficiency. 
This experiment used stable isotopes of Rb with an ionization potential of 4.18 eV. 
The Rb is detected using a Langmuir-Taylor detector~\cite{doi:10.1063/1.1472467}, which consists of a filament that ionizes neutral atoms and a secondary electron multiplier that detects ionized particles. 
The detector cannot detect elements with a high ionization potential because it uses the process of thermal ionization. 
Thus, the sensitivity for detecting Rb was high in this experimental setup because Rb has a small ionization potential and therefore a large ionization efficiency. 

\begin{figure}[!h]
\centering\includegraphics[width=45mm]{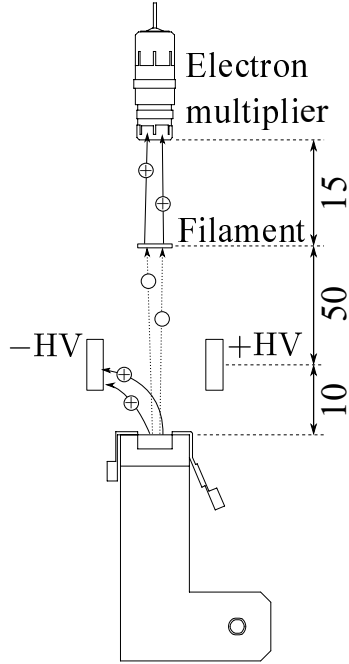}
\caption{
Setup to detect neutralized Rb atoms when the Y foil is positioned upward. 
The neutralized Rb atoms are reionized by the filament and detected by the electron multiplier. 
The sweeper electrode is the same as in Fig.~\ref{Fig:SetupFr}, but the applied voltage is $\pm10$ V. 
When $\pm1000$ V is applied, a stray electric field influences the electron multiplier. 
A preliminary experiment demonstrates that the ions from the foil are completely swept out by $\pm10$ V. 
}
\label{Fig:SetupRb}
\end{figure}

Figure~\ref{Fig:SetupRb} shows the experimental setup used in the following experimental procedure. 
The energy of the Rb ion beam was 900 eV and its current was in the order of nA. 
The beam accumulation time was 30 s and the sequence of the foil heating took 2 s for rising and 10 s for keeping. 
The beam current and accumulation time was close to the detection limit of the Langmuir-Taylor detector. 
Because the ion sweeper was switched on and off repeatedly, the count number was alternately acquired for only neutrals ($\Omega_{\mathrm{atom}} N_{\mathrm{atom}}$) and the total ($\Omega_{\mathrm{ion}} N_{\mathrm{ion}} + \Omega_{\mathrm{atom}} N_{\mathrm{atom}}$). 
Released ions were attracted by the bias voltage of the electron multiplier regardless of the presence of the reionization filament, such that its detection efficiency $\Omega_{\mathrm{ion}}$ would be higher than the efficiency for the neutral $\Omega_{\mathrm{atom}}$. 
Using this procedure, the foil temperature dependence of the neutralization efficiency 
$\rho = (\Omega_{\mathrm{atom}} N_{\mathrm{atom}})/(\Omega_{\mathrm{ion}} N_{\mathrm{ion}} + \Omega_{\mathrm{atom}} N_{\mathrm{atom}} )$
was measured. 
The efficiency $\rho$ is a different parameter from the three parameters derived from the Fr experiment, but it is useful to investigate the relative behavior of the neutralization process. 

As a result, our Rb experiment did not demonstrate reproducible data for the neutralization efficiency. 
The efficiency $\rho$ appeared different every time the foil temperature changed. 
Although the reproducibility was not confirmed, the behavior that the efficiency improved at high temperatures and that the efficiency dropped at low temperatures was observed. 
This behavior might suggest that the Rb that accumulated and stayed on the foil will degrade the efficiency and that the surface condition of the foil will recover after the Rb desorbed at high temperatures, which means that the experimental process itself, including the beam irradiation, accumulation, and heating, affects the condition of the foil surface and consequently affects the effective work function. 
The present experimental result indicates the probability that uncontrollable conditions play important roles in the thermal surface neutralization process. 

\subsection{Fr neutralization with different foils}

For the neutralization of Fr ions, we compared Y foil to Gd ($E_{\mathrm{WF}} = 3.1 \pm 0.15$ eV \cite{PhysRevB.2.1}) and Zr ($E_{\mathrm{WF}}=4.05\pm0.1$ eV \cite{PhysRevB.2.1}) foils. 
Gd and Zr, as well as Y, have previously been used to neutralize radioactive alkali ions \cite{LIPSKI200449, PhysRevLett.86.2967, PhysRevA.97.042507}. 
In 2004, Lipski et al. \cite{LIPSKI200449} measured the residual Fr activity to estimate the release efficiency from Gd foil (and Y and Ba foils). 
In 2001, Crane et al. \cite{PhysRevLett.86.2967} magneto-optically trapped radioactive Rb atoms using Zr foil. 
In 2018, Kalita et al. \cite{PhysRevA.97.042507} magneto-optically trapped radioactive Fr atoms using Zr foil. 
Candidates for neutralizer material are limited because neutralizers should be chemically stable and have a work function less than the ionization potential $E_{\mathrm{IP}}$ of Fr. 
For example, La was initially considered as a candidate because its work function $E_{\mathrm{WF}} = 3.5\pm0.2$ eV \cite{PhysRevB.2.1} and La foil is easy to obtain commercially. 
However, the oxidation of La foil is so strong that the foil broke before the experiment began. 
Additionally, because our neutralizer system uses resistive heating, a low electrical conductivity is preferred. 
We tried to use a 30-$\mu$m-thick Au foil for comparative experiments, but our system was not able to efficiently heat the foil because the electric conductivity of Au is too large. 
Given the laser trapping, a sufficient particle release from a relatively low temperature is preferable. 
For tungsten foil \cite{Melconian200593}, the release begins beyond 1800 K, which is higher than the temperature range for the present work. 
For these reasons, Y, Gd, and Zr were used for this work. 

\begin{figure}[!h]
\centering\includegraphics[width=95mm]{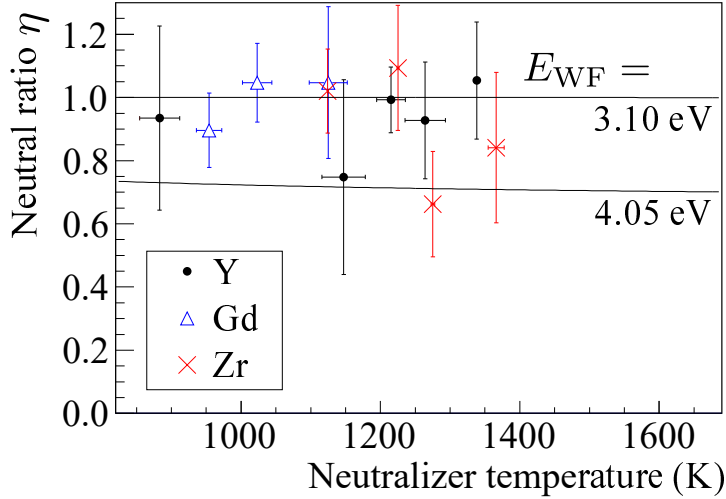}
\caption{
Neutral ratio $\eta$ for Y ($\bullet$), Gd ($\triangle$), and Zr ($\times$) foils. 
Curves are based on the Saha-Langmuir equation with $E_{\mathrm{WF}}=3.10$ eV (for Y and Gd) and 4.05 eV (for Zr) \cite{PhysRevB.2.1}. 
The data for Y are the same as in Fig. \ref{Fig:eta_Y}. 
}
\label{Fig:eta_GdZrY}
\end{figure}

The experimental setup and procedure were similar to those for Fr neutralization with Y foil. 
The Gd and Zr foils were 25 $\mu$m thick (thickness tolerance: 20\%), and each had an effective area of 10 mm $\times$ 10 mm. 
The neutral ratio $\eta$ for different foils is plotted in Fig. \ref{Fig:eta_GdZrY}. 
On the basis of the melting temperature $T_{\mathrm{melt}}$ of the material, significant release begins beyond $T/T_{\mathrm{melt}} \sim 0.5$ and the increase rusts out at $T/T_{\mathrm{melt}} \sim 0.7$ for most materials, according to Ref. \cite{Melconian200593}. 
Therefore, the measurement was carried out in a temperature range around $T/T_{\mathrm{melt}} \sim 0.6$. 
The temperature $T_{\mathrm{melt}}$ for Y, Gd, and Zr is 1799, 1586, and 2128 K, respectively. 
The ratio of SSD count excluding ions and SSD count including ions was determined to obtain $\eta$ as per Eq. (\ref{Eq:etaDef}). 
Because both the numerator and denominator have large statistical errors and because of the large propagated errors, it is difficult to meaningfully differentiate between the foils. 
We hypothesize that a sufficient statistical precision should be obtained if the usual intensity of the Fr beam is supplied, because the intensity in the present experiment was lower than the previously planned intensity. 
The measured efficiency $\eta$ is roughly consistent with the result of the Saha-Langmuir equation (\ref{Eq:SahaLangmuir}) under the assumption that the effective work function does not depart from the ideal work function.

\begin{figure}[!h]
\centering\includegraphics[width=100mm]{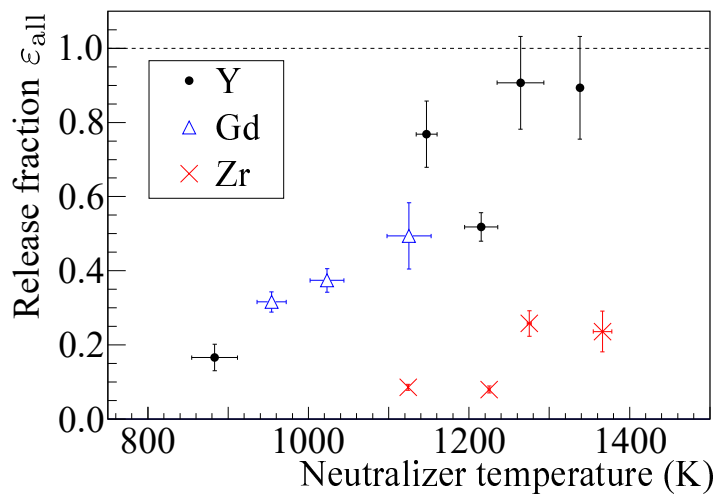}
\caption{
Temperature dependence of the release efficiency for different foils [Y ($\bullet$), Gd ($\triangle$), and Zr ($\times$)]. 
The data for Y are the same as in Fig. \ref{Fig:eps_all_Y}. 
}
\label{Fig:eps_all_GdZrY}
\end{figure}

Figure \ref{Fig:eps_all_GdZrY} shows the measured release efficiency $\varepsilon_{\mathrm{all}}$. 
We confirm that Y foil is superior as a neutralizer target in terms of both the neutral ratio and the release efficiency. 

\section{Conclusion}

The present work demonstrates that thermal surface neutralization, especially when using Y foil, is a more useful neutralization method for Fr MOTs. 
The results reconfirm that Y foil is a good neutralizer (i.e., it has a neutral release efficiency $>65\%$). 
We compared Y, Gd, and Zr foils, and the release fraction $\varepsilon_{\mathrm{all}}$ when using Y foil exceeds 75\% at 1350 K, which is greater than the release fraction for the other foils. 
In this work, the measurements were designed to account for the charge of the released particles, and the measured neutral ratio $\eta$ was close to unity for each foil. 
The neutral ratio $\eta$ is an important parameter for MOTs because it not only leads directly to the number of neutral atoms but also indicates the number of charged particles, which are background components. 
The experimental results show that higher temperatures lead to a more efficient release of particles. 
However, for a MOT experiment, the neutralizing temperature should be optimized in an actual trapping experiment because the trapping efficiency depends strongly on temperature due to the velocity of atoms and the occurrence of background particles. 
In addition, it is clear that the neutralization efficiency depends strongly on experimental conditions. 
Thus, the influence of the ion beam must be considered when interpreting the data. 
Before conducting the Fr MOT experiment, a pilot experiment using a stable Rb isotope is often carried out because the atomic transitions and ionization potential of Rb are close to those of Fr. 
The use of a large amount of Rb could reduce the efficiency of Fr neutralization because the neutralizer target would be strongly influenced by Rb. 
Minimizing such pilot experiments is thus important to maximize the efficiency of Fr neutralization.

\section*{Acknowledgment}

The authors thank Professor M. Itoh and the cyclotron staff at CYRIC. 
This work was supported by JSPS KAKENHI Grants No. 25610112 and No. 26220705. 
The authors would like to thank Enago (www.enago.jp) for the English language review. 

\bibliographystyle{ptephy}

\end{document}